\documentclass[11pt]{article}

\newcommand{\beq}{\begin{eqnarray}}
\newcommand{\eeq}{\end{eqnarray}}
\newcommand{\co}{{\cal{O}}}

\newenvironment{bib}{\begin{list}{}
{\itemindent-7mm \listparindent-7mm
\leftmargin7mm \rightmargin0mm \topsep0mm \partopsep0mm
\setlength{\parsep}{\parskip}}
\item[]}{\end{list}}

\begin{document}

\vspace*{2cm}
\LARGE\bf\baselineskip17pt
ON DETECTING THE GRAVITOMAGNETIC FIELD OF THE EARTH BY MEANS OF ORBITING CLOCKS\\
\\

\normalsize\rm
H.I.M. Lichtenegger$^1$, F. Gronwald$^2$ and B. Mashhoon$^3$\\ \\

$^1${\it Institut f\"ur Weltraumforschung,
         \"Osterreichische Akademie der Wissenschaften,
         A-8010 Graz, Austria},

$^2${\it Institut f\"ur Theoretische Physik,
         Universit\"at zu K\"oln,
         D-50923 K\"oln, Germany},

$^3${\it Department of Physics and Astronomy,
         University of Missouri, Columbia,
         Missouri 65211, USA}\\ \\

ABSTRACT\\

Based on the recent finding that the difference in proper time of
two clocks in prograde and retrograde equatorial orbits about the
Earth is of the order \mbox{$\sim 10^{-7}$~s} per revolution, the possibility of
detecting the terrestrial gravitomagnetic field by means of clocks carried by
satellites is discussed. A mission taking advantage of this influence of the
rotating Earth on the proper time is outlined and the conceptual difficulties are
briefly examined.\\ \\

INTRODUCTION\\

General relativity was designed to be a field theory formally similar 
to electrodynamics, yet some basic differences remain. However,
in the case of low velocities and weak fields, the formal analogy 
between gravity and electromagnetism gives rise to a number of similar 
phenomena, known as gravitoelectromagnetism. Due to the weakness of the 
gravitomagnetic effects, however, their existence has yet to
be verified.\\

The proof whether rotating masses do indeed modify the gravitational field is
important, both from a physical as well as a philosophical point of view, since it
is intimately related to our view of space as being either relational or absolute.
Hitherto discussed solely on a philosophical level, Newton raised the problem of
the status of space on a physical basis by ascribing to absolute space a physical
reality which becomes apparent via inertial forces. By criticising Newton's concept
of abolute space, Mach stated that dynamics should not contain absolute elements
since in his positivistic view observable effects should have observable causes.
Because inertial forces are observed only to occur in reference frames accelerated
with respect to the fixed stars, for Mach it was therefore obvious to suppose that
the distribution of matter in the universe determines the local inertial frames and
the reliance on absolute space becomes superfluous (Mach's principle). Although it
is still a matter of debate to what extent general relativity complies with Mach's
principle (see e.g. Mashhoon, 1988, and Barbour and Pfister, 1995), it is generally
believed that certain gravitomagnetic effects are the most direct manifestations of
Machian ideas in general relativity (Mashhoon, 1993). Hence the study of these
phenomena can help to clarify one of the most puzzling questions in physics: Where
does inertia come from?\\

Several space missions have been proposed to measure gravitomagnetism directly,
among them Gravity Probe-B, LAGEOS III and the Superconducting Gravity Gradiometer
Mission (see e.g. Ciufolini and Wheeler, 1995). The Gravity Probe-B satellite will
carry four superconducting spherical quartz gyroscopes to measure the precession of
the gyroscopes relative to the distant stars. LAGEOS III could be launched in an
orbit whose inclination is supplementary to that of LAGEOS I and to measure the
precession of the line of nodes of these two satellites. The Superconducting
Gravity Gradiometer Mission envisages orbiting an array of three mutually
orthogonal superconducting gravity gradiometers around the Earth to measure
directly the contribution of the gravitomagnetic field to the tidal gravitational
force.\\

In contrast to the above proposals, in this paper we consider the possibility to
detect gravitomagnetism by means of clocks, i.e. by measuring the difference in
proper time displayed by two clocks orbiting in pro- and retrograde direction
around the Earth.\\

THE GRAVITOMAGNETIC CLOCK EFFECT\\

The gravitational field outside a rotating body of mass $M$ and angular momentum
$J$ is given by the Kerr metric \beq\label{kerr} ds^2 =
\left(1-\frac{R_s\,r}{\rho^2}\right)c^2dt^2+
         \frac{2R_s\,r}{\rho^2}a\sin^2\theta\, cdt\,d\phi-
         {{\rho^2}\over{\Delta}}dr^2-\rho^2 d\theta^2-
         \sin^2\theta\left[r^2+a^2+\frac{R_s\,r}{\rho^2}a^2\sin^2\theta\right]d\phi
         ^2,
\eeq
with
\[ \rho^2=r^2+a^2\cos^2\theta, \hspace{6mm} \Delta = r^2-R_s r+a^2 \]
and
\[ a=\frac{J}{cM}, \hspace{6mm}R_s=\frac{2GM}{c^2}.\]
Restricting ourselves to a circular equatorial orbit, one of the geodesic equations
of Eq. (\ref{kerr}) becomes
\beq\label{kerrgeo}
\frac{d\phi}{dt}\equiv\Omega=
   \Omega_0\left[\pm
   1-\frac{a}{c}\Omega_0+\co\left(\frac{a^2}{c^2}\right)\right]\cong
   \,\pm\Omega_0+\Omega_{\rm\small LT},
\eeq
where the gravitomagnetic correction in the equatorial plane to the Kepler period
$\Omega_0=(GM/r^3)^\frac{1}{2}$ (as determined by static observers at infinity) is
just the Lense-Thirring precession $\Omega_{\rm LT}=-G\,J/c^2r^3$. Now we consider
two clocks moving along pro- and retrogarde circular equatorial orbits,
respectively, about the Earth. According to Eq. (\ref{kerrgeo}), the difference in
proper time for the two clocks after some fixed coordinate time $t$, say, one
Kepler period $t=T_0=2\pi/\Omega_0$ becomes ($\tau_+$ and $\tau_-$ denote the
proper time along the pro- and retrograde orbits, respectively)
\beq\label{deltataut} (\tau_+-\tau_-)_{t=T_0} \cong 12\pi\frac{GJ}{c^4 r}\approx
 3\times 10^{-16}\, {\rm s},  
\eeq
where we have inserted the values $M=6\times 10^{24}\,{\rm kg},\:
J=6\times 10^{33}\,{\rm kg\,m^2\,s^{-1}}$ and $r=7\,000$\, km. This is a very tiny
effect which, even apart from all perturbations, is not detectable with today's
atomic clock technology and which is probably the reason why clocks are barely
considered as proper means to measure the gravitomagnetic contribution of the
Earth's rotation. However, as recently pointed out by Cohen and Mashhoon (1993),
this time difference is considerably enlarged when calculating the difference in
proper time of two counter-orbiting clocks for some fixed angular interval rather
than for some fixed time interval. Integration over, say, $2\pi$ yields (see
Gronwald {\it et al}., 1997)
\beq\label{deltatauphi}
(\tau_+-\tau_-)_{\phi=2\pi}\cong4\pi\frac{J}{Mc^2}\approx 1\times 10^{-7}\,{\rm s}
\eeq
and it is interesting to note that this difference is independent of both $G$
and $r$. The relatively large value in Eq. (\ref{deltatauphi}) could underlie an
experiment to directly measure the terrestrial gravitomagnetic field (called
Gravity Probe-C (GP-C) in Gronwald {\it et al}. (1997)). The above analysis can be
extented to circular orbits with finite inclination (Theiss, 1985), showing that
the time difference decreases with increasing inclination; for a polar orbit the
effect vanishes as it is expected because of symmetry.\\

ERROR SOURCES\\

Although the value in Eq. (\ref{deltatauphi}) is several orders of magnitude larger
than that in Eq. (\ref{deltataut}) and well within the measurement capabilities of
modern atomic clocks ($(\tau_+-\tau_-)/T_0\sim 10^{-11}$ for a near-Earth orbit),
it is not a simple task to seperate the time difference induced by gravitomagnetism
from all other effects resulting likewise in a time shift between the clocks. The
main error sources that may affect the experiment can be divided into 4 categories
and are listed below:
\begin{enumerate}
\item Errors from orbital injection errors

\item Errors from gravitational perturbations
\begin{itemize}
 \item[-] Static odd zonal harmonic perturbations
 \item[-] Static non-zonal harmonic perturbations
 \item[-] Non-linear harmonic perturbations
 \item[-] Solid and ocean Earth tides
 \item[-] Sun, Moon and planetary tidal accelerations
 \item[-] Relativistic perturbations
\end{itemize}
\item Errors from non-gravitational perturbations
\begin{itemize}
 \item[-] Direct solar radiation pressure
 \item[-] Terrestrial radiation pressure (albedo, infrared radiation)
 \item[-] Satellite eclipses
 \item[-] Anisotropic thermal radiation (Yarkovsky effect)
 \item[-] Thermal thrust (Yarkovsky-Schach effect)
 \item[-] Poynting-Robertson effect
 \item[-] Atmospheric drag
 \item[-] Drag from interplanetary dust
 \item[-] Earth's plasma environment (charged particle drag)
 \item[-] Earth's magnetic field (magnetic despin)
\end{itemize}
\item Errors from uncertainties in the determination of the orbital parameters
\end{enumerate} After launch, the orbital elements have to be adjusted by inflight
corrections as close as possible to the values desired, i.e. to zero eccentricity
and inclination in order to maximize the effect. Further, an accurate tracking of
the satellites is required since uncertainties in the radial and azimuthal location
of the clocks readily cover the gravitomagnetic effect. A simple error analysis
yields $\Delta r/r \sim \Delta\phi/\phi\sim\Omega_0 J/M c^2\sim 10^{-11}$, i.e. one
must keep track of the satellites with an accuracy of $\Delta r\leq 0.1$ mm and
$\Delta\phi\leq 10^{-2}$ mas per revolution for orbits with $\sim 7\,000$\, km
altitude (Gronwald {\it et al}., 1997). Equivalently, accelerations down to $\Delta
r/T_0^2\sim 10^{-11}$ g should be taken into account.\\

Gravitational perturbations due to the nonsphericity of the Earth will at least
partially cancel out for two satellites following opposite orbits. While current
models of the static and tidal Earth's gravitational field induce orbital errors in
all three spatial directions of several mm (Ciufolini {\it et al}., 1998), the
gravitational influence of other celestial bodies on the satellites is essentialy
due to the Moon ($10^{-7}$~g) and the Sun ($10^{-8}$~g) and can be modeled with
sufficient accuracy (Gronwald {\it et al}., 1997). Finally, accelerations due to
relativistic effects other than the relativistic Kerr corrections to the
terrestrial field can be neglected (Mashhoon and Theiss, 1982, Ashby and Bertotti,
1984).\\

Several non-gravitational mechanisms which perturb the orbits of satellites have
been modeled in some detail in the course of data analysis of laser-ranged
satellites (e.g. Rubincam, 1982). The observed average drag on the LAGEOS satellite
is $\sim 10^{-12}\,{\rm m/s^2}$ and is due to Yarkovsky thermal, neutral and
charged particle drag (Rubincam, 1990); thermal thrust induces a
secular variation in inclination of LAGEOS of the order of $\sim 1$\,mas/yr
(Farinella {\it et al}., 1990). Although the perturbations will depend on
inclination and are also controlled by the eclipses (the LAGEOS satellites are on
nearly polar orbits), non-gravitational perturbations on LAGEOS-type satellites
(i.e. spherically symmetric, small ratio of cross-sectional area to mass
satellites) along equatorial orbits are not expected to differ significantly from
those in polar orbits. Based on the experience of LAGEOS's orbital variations, it
might thus be possible to correct for the non-gravitational perturbations to the
desired order without the use of drag-free satellites; however, a conclusive answer
cannot be given without a thorough analysis of these effects.\\

SUMMARY AND CONCLUSION\\

Based on the fact that the difference in proper time of two counter-orbiting clocks
along circular equatorial orbits after completing one revolution each is several
orders of magnitude larger than the time difference after one Kepler period, a
mission is proposed to directly measure the gravitomagnetic field of the Earth via
spaceborne atomic clocks. While many of the effects causing deviations from an
ideal circular Keplerian orbit can be incorporated into the analysis, the accurate
tracking of the satellites and the correct modeling of the dynamical part of the
Earth gravity field seem to be the most difficult tasks for a Gravity Probe-C
mission. However, the reported clock effect is cumulative and after 100-1000
revolutions it should be detectable even with today's tracking techniques. Also,
the next generation of static and time-dependent terrestrial gravity field models
is expected to possess a precision sufficient for GP-C. Hence we consider GP-C as
an alternative way to detect gravitomagnetism, which ideally completes missions
like Gravity Probe-B and appears technically less demanding.\\

REFERENCES

\begin{bib}
{Ashby, N.} and B.~Bertotti,
\newblock Relativistic Perturbations of an Earth Satellite,
\newblock {\em Phys. Rev. Lett.}, {\bf 52}, 485--488 (1984).

{Barbour, J.} and H.~Pfister, editors,
\newblock {\em {Mach's Principle: From Newton's Bucket to Quantum Gravity}},
  volume~6 of {\em Einstein Studies},
\newblock Birkh\"auser (1995).

{Ciufolini, I.} and J.A. Wheeler,
\newblock {\em {Gravitation and Inertia}},
\newblock Princeton Series in Physics, Princeton (1995).

{Ciufolini, I.}, E.~Pavlis, F.~Chieppa, E.~Fernandes-Vieira, and
  J.~Perez-Mercader,
\newblock Test of General Relativity and Measurment of the {L}ense-{T}hirring
  Effect with Two Earth Satellites,
\newblock {\em Science}, {\bf 279}, 2100--2103 (1998).

{Cohen, J.M.} and B.~Mashhoon,
\newblock Standard Clocks, Interferometry, and Gravitomagnetism,
\newblock {\em Phys.\,Lett.\,A}, {\bf 181}, 353--358 (1993).

{Farinella, P.}, A.M. Nobili, F.~Barlier, and F.~Mignard,
\newblock Effects of Thermal Thrust on the Node and Inclination of {L}ageos
\newblock {\em Astron.\,Astrophys.}, {\bf 234}, 546--554 (1990).

{Gronwald, F.}, E.~Gruber, H.~Lichtenegger, and R.A. Puntigam,
\newblock Gravity Probe {C{\tiny lock}} - {P}robing the Gravitomagnetic Field
  of the {E}arth by Means of a Clock Experiment,
\newblock in {\em Fundamental Physics in Space}, pp. 29--37, ESA SP-420 (1997).

{Mashhoon, B.} and D.S. Theiss,
\newblock Relativistic Tidal Forces and the Possibility of Measuring Them, 
\newblock {\em Phys. Rev. Lett.}, {\bf 49}, 1542--1545 (1982). 

{Mashhoon, B.},
\newblock Complementarity of Absolute and Relative Motion,
\newblock {\em Phys. Lett. A}, {\bf 126}, 393--399 (1988).      

{Mashhoon, B.},
\newblock On the Relativity of Rotation, 
\newblock in {\em Directions in General Relativity}, vol. 2, edited by
 B.~L. Hu and T.~A. Jacobson, pp. 182--194, Cambridge University Press,
 Cambridge (1993).

{Rubincam, D.P.},
\newblock On the Secular Decrease in the Semimajor Axis of {LAGEOS's} Orbit,
\newblock {\em Celest.\,Mech.}, {\bf 26}, 361--382 (1982).

{Rubincam, D.P.},
\newblock Drag on the {LAGEOS} Satellite,
\newblock {\em J.\,Geophys.\,Res.}, {\bf 95}, 4881--4886 (1990).

{Theiss, D.S.},
\newblock A General Relativistic Effect of a Rotating Spherical Mass and the
  Possibility of Measuring it in a Space Experiment,
\newblock {\em Phys.\,Lett.\,A}, {\bf 109}, 19--22 (1985).

\end{bib}

\end{document}